\documentclass[a4paper,fleqn,usenatbib]{mnras}

\usepackage{newtxtext,newtxmath}
\usepackage[T1]{fontenc}
\usepackage{ae,aecompl}

\usepackage{graphicx}	% Including figure files
\usepackage{amsmath}	% Advanced maths commands
\usepackage{amssymb}	% Extra maths symbols

\def\teff{$T_{\text {eff}}$}
\def\vt{$V_{\text {t}}$}
\def\logg{$\log\,g$}
\def\v0{$V_{0}$}
\def\vsini{$V$sin$i$}

\title[Oxygen abundance for AFG supergiants]{Oxygen abundance and the N/C vs N/O relation for AFG supergiants and bright giants}

\author[L.S.~Lyubimkov et al.]{
L.S.~Lyubimkov,$^{1}$\thanks{E-mail: lyub@craocrimea.ru}
S.A.~Korotin,$^{1}$
and D.L.~Lambert$^{2}$
\\
$^{1}$Crimean Astrophysical Observatory, Nauchny 298409, Republic of Crimea\\
$^{2}$W.J. McDonald Observatory and Department of Astronomy, The University of Texas at Austin, Austin, TX 78712, USA\\
}

\date{Accepted XXX. Received YYY; in original form ZZZ}

\pubyear{2019}

\begin{document}
\label{firstpage}
\pagerange{\pageref{firstpage}--\pageref{lastpage}}
\maketitle

\begin{abstract}

Non-LTE analysis (LTE is local thermodynamic equilibrium) of the oxygen 
abundances for 51 Galactic A-, F- and G-type supergiants and bright giants 
is performed. In contrast with carbon and nitrogen, oxygen does not show any 
significant systematic anomalies in their abundances log $\varepsilon$(O). 
There is no marked difference from the initial oxygen abundance within errors 
of the log $\varepsilon$(O) determination across the \teff\ interval from 4500 
to 8500 K and the \logg\ interval from 1.2 to 2.9 dex. This result agrees well 
with theoretical predictions for stellar models with rotation. With our new 
data for oxygen and our earlier non-LTE determinations of the N and C 
abundances for stars from the same sample, we constructed the [N/C] vs [N/O] 
relation for 17 stars. This relation is known to be a sensitive indicator of 
stellar evolution. A pronounced correlation between [N/C] vs [N/O] is found; 
the observed [N/C] increase from 0 to 1.6 dex is accompanied by the [N/O] 
increase from 0 to 0.9 dex. When comparing the observed [N/C] vs [N/O] relation 
with the theoretical one, we show that this relation reflects a strong 
dependence of the evolutionary changes in CNO abundances on the initial 
rotation velocities of stars. Given that the initial rotational velocities of 
these stars are expected to satisfy \v0$<$150 km/s, it is found that they are 
mostly the post first dredge-up (post-FDU) objects. It is important that just 
such initial velocities \v0 are typical for about 80 \% of stars in question 
(i.e. for stars with masses 4-19 M$_{\sun}$). A constancy of the total C+N+O 
abundance during stellar evolution is confirmed. The mean value 
log $\varepsilon$(C+N+O)=8.97$\pm$0.08 found for AFG supergiants and bright 
giants seems to be very close to the initial value 8.92 (the Sun) or 8.94 
(the unevolved B-type MS stars).

\end{abstract}


%%%%%%%%%%%%%%%%%%%%%%%%%%%%%%%%%%%%%%%%%%%%%%%%%%

%%%%%%%%%%%%%%%%% BODY OF PAPER %%%%%%%%%%%%%%%%%%

\section{Introduction}

Carbon, nitrogen and oxygen are key light chemical elements whose atmospheric 
abundances may significantly change during the stellar evolution. Therefore, a 
study of these elements provides an opportunity for testing modern evolutionary
theories. We consider here stars with masses $M$ between 4 and 20 M$_{\sun}$, 
which are observed at first as early and middle B-type stars on the Main 
Sequence (MS), i.e., luminosity classes V, IV and III. As is well known, 
the H-burning CNO-cycle is the main source of energy in these stars with such $M$ during this 
evolutionary phase. Next, these stars are observed as A, F and G-type 
supergiants and bright giants, i.e., luminosity classes I and II. These luminous stars 
are the focus of the present paper. 

Rotation of a star plays an important role in evolution of stars with masses $M$ = 
4-20 M$_{\sun}$. Rotation affects the stellar interior and may also affect the surface chemical composition. When the initial rotational velocity 
\v0 is rather high, the rotationally-induced mixing can lead to marked changes 
in the surface C and N abundances already during the MS stage. Further 
alterations in the surface C and N abundances can take place on the next stage 
of A, F and G supergiants, especially after the deep convecting mixing, the 
so-called First Dredge-Up (FDU). Since the surface and interior abundances of MS stars may depend on the stellar rotation so are the surface abundances in post-FDU dependent on the stellar rotation. The stars in their post-MS phase and after the
FDU (post- FDU phase) are of special interest to us. 

Our sample of  the Galactic A-, F- and G-type supergiants and bright giants has 
been studied by us earlier in a series of papers providing: (a) accurate 
determination of fundamental parameters for 63 stars, including their effective 
temperature \teff, surface gravity \logg, microturbulent velocity \vt~ and index 
of metallicity [Fe/H] (\citealt{Lyubimkov10}, Paper I); (b) the non-LTE 
(LTE is local thermodynamic equilibrium) analysis
of the nitrogen abundance for 30 stars that confirmed the N enrichment in 
atmospheres of this type stars (\citealt{Lyubimkov11}, Paper II); (c) the 
non-LTE analysis of the lithium abundance for 55 stars which led to some 
interesting conclusions after comparison with theory 
(\citealt{Lyubimkov12}, Paper III); (d) the non-LTE analysis of the carbon 
abundance for 36 stars that confirmed both the expected C general deficiency in their 
atmospheres and the N vs C anticorrelation (\citealt{Lyubimkov15}, Paper IV). 

Now, we present our results of the non-LTE oxygen abundance determination for 51 
A-, F- and G-type supergiants and bright giants. We used in this analysis the 
parameters \teff, \logg, \vt~ and [Fe/H] found in Paper I. These new data for 
oxygen together with our earlier results for nitrogen (Paper II) and carbon 
(Paper IV) allow us construct the N/C vs N/O relation for 17 stars. 
According to \citet{Maeder14}, the N/C vs N/O relation is a very sensitive 
indicator of mixing in stars. These authors noted that the N/C vs N/O plot can 
be considered as ''an ideal quality test for observational results``. 
%We may add that at the same time this relation is an important test for 
%theoretical results. 

There are a lot of works on the C, N and O abundance determination for 
late-type stars, mostly for dwarfs and giants. However, data on the CNO 
abundances for FGK supergiants are rather limited, especially as for the O 
abundance. On the one hand, it was found with confidence, at first in early LTE 
determinations (see, e.g., \citealt{Luck78}; \citealt{Luck85} and later in 
non-LTE analyses (see, e.g., our Papers II and IV), that the systematic C 
underabundance and the N overabundance (as compared with the Sun) take place 
in these stars. 

On the other hand, not numerous data on the O abundance were less definite. 
In particular, a mild oxygen deficiency was found for FGK supergiants in LTE 
analysis of \citet{Luck85}. A question arose: is oxygen really 
somewhat underabundant in such stars? The later non-LTE determinations did not 
touch upon this problem. For instance, in the work of \citet{Korotin14} 
a non-LTE analysis of the O abundances was implemented for a large sample of 
variable supergiants (cepheids) with various distances. It was found there 
that marked variations in the derived O abundances are connected with 
different Galactocentric distances of the stars (they vary from 5 to 17 kpc). 
An aim of this work was a determination of the radial O/H gradient in Galactic 
thin disk.  

The aim of our work is a non-LTE determination of O abundances for non-variable
AFG supergiants and bright giants which are mostly within 1 kpc from the Sun. 
The derived values for these nearby stars are compared with the initial O 
abundance; their dependence on basic parameters \teff\  and \logg\ is considered 
as well. Both the obtained O abundances and the constructed N/C vs N/O relation 
are compared with theoretical predictions based on stellar models with rotation.
The total C+N+O abundance for these evolved stars is determined; its 
conservation during stellar evolution is discussed.

%There are a lot of works on C, N and O abundance determinations for 
%late-type stars. They concern primarily the LTE-analysis of the CNO abundances. A review of these works is outside our paper. Here we analyze the non-LTE abundances for all three elements in stars on a very interesting stage of evolution, where the majority but perhaps not all have passed through the FDU.

\section{SPECTRAL OBSERVATIONS AND THE \ion{O}{i} LINE LIST}

We base the present work, like Papers I-IV, on the high-resolution spectral 
observations of Galactic AFG-type supergiants and bright giants, which have 
been acquired by us at the W.J. McDonald Observatory in 2003-2006 using the 2.7-m 
telescope and the Tull echelle spectrometer \citep{Tull95}. Additional 
observations of some stars were performed in 2009 with the same spectrometer. 
The stars were observed at a resolving power of R = 60000. The spectral region 
from about 4000 to 9700 \AA\ was covered. The typical signal-to-noise ratio of 
the extracted one-dimensional spectra was between 100 and 400. Two spectra for 
each programme star have been obtained. 

The list of \ion{O}{i} lines used in our non-LTE analysis is presented in 
Table \ref{line}. The oscillator strengths log $gf$ here are taken from the 
NIST database \citep{NIST}
%\footnote{\url{https://physics.nist.gov/asd}} 
and the damping constants 
$\gamma_{\text {rad}}$, $\gamma_{\text {st}}$ and $\gamma_{\text {vdw}}$ are taken from the VALD 
database \citep{Ryabchikova15}.

\begin{table}
%\centering
\caption{List of the used \ion{O}{i} lines.}
\label{line}
\begin{tabular}{lrrrcc}
\hline
$\lambda$ (\AA)& E$_{\rm low}$(eV)& log $gf$& $\gamma_{\text {rad}}$&$\gamma_{\text {st}}$&$\gamma_{\text {vdw}}$\\
\hline
\hline
 5577.34&  1.96& -8.20&    0.15&    -6.29&   -8.04\\
 6155.96& 10.74& -1.36&    7.62&    -3.95&   -6.85\\
 6155.97& 10.74& -1.01&    7.62&    -3.95&   -6.85\\
 6155.99& 10.74& -1.12&    7.61&    -3.95&   -6.85\\
 6156.74& 10.74& -1.49&    7.62&    -3.95&   -6.85\\
 6156.76& 10.74& -0.90&    7.62&    -3.95&   -6.85\\
 6156.78& 10.74& -0.69&    7.61&    -3.95&   -6.85\\
 6158.15& 10.74& -1.84&    7.62&    -3.95&   -6.85\\
 6158.18& 10.74& -1.00&    7.62&    -3.95&   -6.85\\
 6158.19& 10.74& -0.41&    7.61&    -3.95&   -6.85\\
 6300.30&  0.00& -9.72&   -2.16&    -8.12&   -6.48\\
 7771.94&  9.15&  0.37&    7.52&    -5.54&   -7.47\\
 7774.16&  9.15&  0.22&    7.52&    -5.54&   -7.47\\
 7775.39&  9.15&  0.00&    7.52&    -5.54&   -7.47\\
 8446.25&  9.52& -0.46&    8.77&    -5.43&   -7.60\\
 8446.36&  9.52&  0.24&    8.77&    -5.43&   -7.60\\
 8446.76&  9.52&  0.01&    8.77&    -5.43&   -7.60\\
 9260.81& 10.74& -0.24&    7.87&    -4.94&   -7.42\\
 9260.85& 10.74&  0.11&    7.88&    -4.94&   -7.42\\
 9260.94& 10.74&  0.00&    7.90&    -4.94&   -7.42\\
 9262.58& 10.74& -0.37&    7.88&    -4.94&   -7.42\\
 9262.67& 10.74&  0.22&    7.90&    -4.94&   -7.42\\
 9262.78& 10.74&  0.43&    7.93&    -4.94&   -7.42\\
 9265.83& 10.74& -0.72&    7.90&    -4.94&   -7.42\\
 9265.93& 10.74&  0.13&    7.94&    -4.94&   -7.42\\
 9266.01& 10.74&  0.71&    7.88&    -4.94&   -7.42\\
\hline
\end{tabular}
\end{table}

Three infrared triplets are included in Table \ref{line}, namely 7771-7775 \AA, 8446 \AA\
and 9260-9266 \AA. Large non-LTE effects take place for these lines, which lead
to their significant strengthening (see below). Unfortunately, the 9260-9266 \AA\
triplet is often seriously blended with telluric lines, so it was of limited use. Apart 
from the generally strong IR lines, we also considered  the following generally 
weak lines in the visual spectral region: the forbidden 5577 \AA\ and 6300 \AA\ 
lines and the permitted lines at 6155, 6156 and 6158 \AA.

%\section{SELECTION OF STARS}

%We determined the non-LTE oxygen abundance for the 51 stars; their list is 
%presented in Table \ref{star}. We included in this list the stars from Paper I for 
%which \ion{O}{i} lines were observable. However, some of these stars were 
%excluded from final list because the O abundances derived for them from 
%\ion{O}{i} lines in IR and visual regions showed a marked discrepancy
%(about 0.3 dex or more). For instance, such a discrepancy 
%was obtained for the two hottest A5-type supergiants, HR~825 (\teff = 8570~K) 
%and HR~2874 (\teff = 8620~K). It should be noted that non-LTE corrections in O 
%abundances for IR lines increase with \teff, so for these relatively hot stars 
%they are especially great. Apart from the possible uncertainties in the \teff\ 
%values for these supergiants, errors in their high microturbulent parameters, 
%\vt = 10.8 and 7.8 km/s, can distort the results from the strong IR triplets. 
%It should be noted as well that a possible dependence of \vt\ on height in 
%atmospheres of such luminous stars can play a marked role. Such dependence was 
%found, for example, for F supergiants by \citet{Lyubimkov90}. 

\begin{table*}
\centering
\caption{List of 51 programme stars with their parameters and the derived 
oxygen abundances.}
\label{star}
\begin{tabular}{llccccrc}
\hline
HR & Sp& \teff & \logg& \vt (km/s)& M/M$_{\sun}$&  [Fe/H]& log $\varepsilon$(O)\\
\hline
\hline
  27&  F2 II     & 6270 &2.10&  3.6  &   6.1  &   -0.09 & 8.97$\pm$  0.20\\
 207&  G0 Ib     & 5220 &1.55&  4.0  &   7.9  &   -0.12 & 8.76$\pm$  0.15\\
 292&  F0 II     & 6880 &2.05&  2.7  &   7.1  &    0.05 & 8.89$\pm$  0.18\\
 461&  G5 II     & 4430 &1.18&  2.8  &   9.5  &   -0.03 & 8.69$\pm$  0.20\\
 792&  G5 II     & 5020 &2.09&  2.9  &   5.0  &   -0.01 & 8.72$\pm$  0.10\\
 849&  G5 Iab    & 5020 &1.73&  2.2  &   6.5  &    0.08 & 8.77$\pm$  0.10\\
1017&  F5 Ib     & 6350 &1.90&  5.3  &   7.3  &   -0.07 & 8.84$\pm$  0.10\\
1135&  F5 II     & 6560 &2.44&  3.5  &   4.8  &    0.06 & 8.74$\pm$  0.20\\
1270&  G8 IIa    & 5060 &1.91&  1.7  &   5.7  &    0.09 & 8.68$\pm$  0.10\\
1303&  G0 Ib     & 5380 &1.73&  3.6  &   7.0  &   -0.09 & 8.67$\pm$  0.10\\
1327&  G5 IIb    & 5440 &2.89&  1.2  &   2.8  &   -0.09 & 8.69$\pm$  0.12\\
1603&  Ib-IIa    & 5300 &1.79&  4.8  &   6.5  &   -0.04 & 8.81$\pm$  0.18\\
1740&  A5 II     & 8300 &2.10&  4.3  &   8.8  &   -0.08 & 8.70$\pm$  0.15\\
1829&  G5 II     & 5450 &2.60&  1.3  &   3.5  &   -0.09 & 8.61$\pm$  0.15\\
1865&  F0 Ib     & 6850 &1.34&  3.9  &  13.9  &    0.03 & 8.74$\pm$  0.15\\
2000&  G2 Ib-II  & 5000 &2.45&  2.1  &   3.9  &   -0.10 & 8.69$\pm$  0.15\\
2453&  G5 Ib     & 4900 &1.70&  2.3  &   6.5  &    0.12 & 8.76$\pm$  0.15\\
2597&  F2 Ib-II  & 6710 &2.02&  3.3  &   7.1  &   -0.18 & 8.79$\pm$  0.20\\
2786&  G2 Ib     & 5260 &1.90&  3.2  &   5.9  &    0.05 & 8.77$\pm$  0.12\\
2833&  G3 Ib     & 5380 &2.21&  4.0  &   4.7  &    0.08 & 8.81$\pm$  0.13\\
2881&  G3 Ib     & 5300 &1.66&  5.2  &   7.3  &   -0.17 & 8.66$\pm$  0.15\\
3045&  G6 Iab-Ib & 4880 &1.21&  5.1  &   9.9  &   -0.07 & 8.71$\pm$  0.18\\
3073&  F1 Ia     & 6670 &2.61&  3.5  &   4.2  &    0.10 & 8.84$\pm$  0.20\\
3102&  F7 II     & 5690 &2.17&  3.7  &   5.1  &    0.11 & 8.87$\pm$  0.20\\
3183&  A 5 II    & 8530 &2.67&  3.5  &   5.4  &    0.04 & 8.75$\pm$  0.18\\
3188&  G2 Ib     & 5210 &1.75&  3.3  &   6.6  &    0.01 & 8.75$\pm$  0.12\\
3229&  G5 II     & 5130 &2.04&  2.3  &   5.2  &    0.01 & 8.73$\pm$  0.15\\
3459&  G1 Ib     & 5370 &2.08&  3.5  &   5.2  &    0.03 & 8.73$\pm$  0.12\\
4166&  G2.5 IIa  & 5475 &2.36&  2.7  &   4.2  &    0.02 & 8.73$\pm$  0.12\\
4786&  G5 II     & 5100 &2.52&  1.5  &   3.7  &    0.10 & 8.78$\pm$  0.12\\
5165&  G0 Ib-IIa & 5430 &2.37&  2.5  &   4.2  &   -0.17 & 8.82$\pm$  0.15\\
6081&  A5 II     & 8370 &2.12&  2.8  &   8.7  &    0.03 & 8.67$\pm$  0.15\\
6536&  G2 Ib-IIa & 5160 &1.93&  3.0  &   6.0  &    0.02 & 8.78$\pm$  0.15\\
6978&  F7 Ib     & 6000 &1.70&  4.6  &   8.2  &   -0.09 & 8.74$\pm$  0.15\\
7014&  F2 Ib     & 6760 &1.66&  4.6  &  10.0  &   -0.07 & 8.89$\pm$  0.20\\
7094&  F2 Ib     & 6730 &1.75&  3.4  &   9.1  &   -0.19 & 8.61$\pm$  0.18\\
7164&  G3 II     & 5200 &2.25&  2.5  &   4.5  &   -0.10 & 8.71$\pm$  0.12\\
7264&  F2 II     & 6590 &2.21&  3.2  &   5.9  &   -0.17 & 8.71$\pm$  0.20\\
7387&  F3 Ib     & 6700 &1.43&  4.4  &  12.5  &   -0.03 & 8.71$\pm$  0.20\\
7456&  G0 Ib     & 5550 &2.06&  2.8  &   5.5  &   -0.16 & 8.63$\pm$  0.10\\
7542&  F8 Ib-II  & 5750 &2.15&  4.2  &   5.3  &    0.17 & 8.79$\pm$  0.18\\
7770&  F5 Ib     & 6180 &1.53&  5.0  &  10.0  &   -0.22 & 8.64$\pm$  0.15\\
7795&  G5 III+A  & 4870 &2.00&  3.1  &   5.3  &    0.00 & 8.72$\pm$  0.10\\
7823&  F1 II     & 6760 &1.92&  4.2  &   7.8  &   -0.13 & 8.71$\pm$  0.15\\
7834&  F5 II     & 6570 &2.32&  3.6  &   5.3  &    0.00 & 8.84$\pm$  0.14\\
7876&  A9 II     & 7020 &1.66&  3.4  &  10.6  &   -0.21 & 8.76$\pm$  0.20\\
8232&  G0 Ib     & 5490 &1.86&  3.7  &   6.4  &    0.10 & 8.73$\pm$  0.17\\
8313&  G5 Ib     & 4910 &1.58&  2.8  &   7.1  &   -0.02 & 8.65$\pm$  0.18\\
8412&  G5 Ia     & 5280 &2.35&  2.3  &   4.2  &    0.05 & 8.72$\pm$  0.18\\
8414&  G2 Ib     & 5210 &1.76&  3.8  &   6.5  &    0.03 & 8.77$\pm$  0.15\\
8692&  G4 Ib     & 4960 &1.90&  3.4  &   5.6  &   -0.10 & 8.68$\pm$  0.15\\
\hline
\end{tabular}
\end{table*}

%There are also several cooler stars, for which we cannot find the O abundance 
%with confidence. A marked discrepancy between \ion{O}{i} lines in IR and visual 
%regions was obtained for some of them, too. Possible errors in the parameters \teff\ and 
%\vt\ can be a reason of the discrepancy. It should be noted that our results for 
%oxygen on the whole seem to be more sensitive to the \teff\ and \vt\ variations then 
%the previous results for carbon and nitrogen. We excluded these stars from 
%Table \ref{star}, however we hope to redetermine in future more exactly both their basic 
%parameters and the CNO abundances. 

\section{NON-LTE COMPUTATIONS OF \ion{O}{i} LINES}

\begin{figure*}
\includegraphics[width=18cm,clip=true]{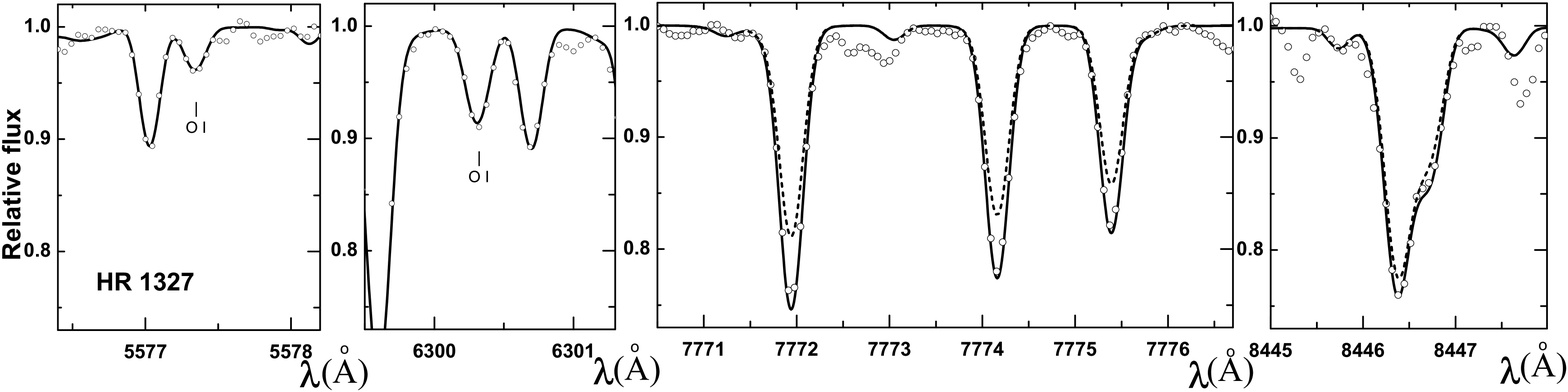}
\includegraphics[width=18cm,clip=true]{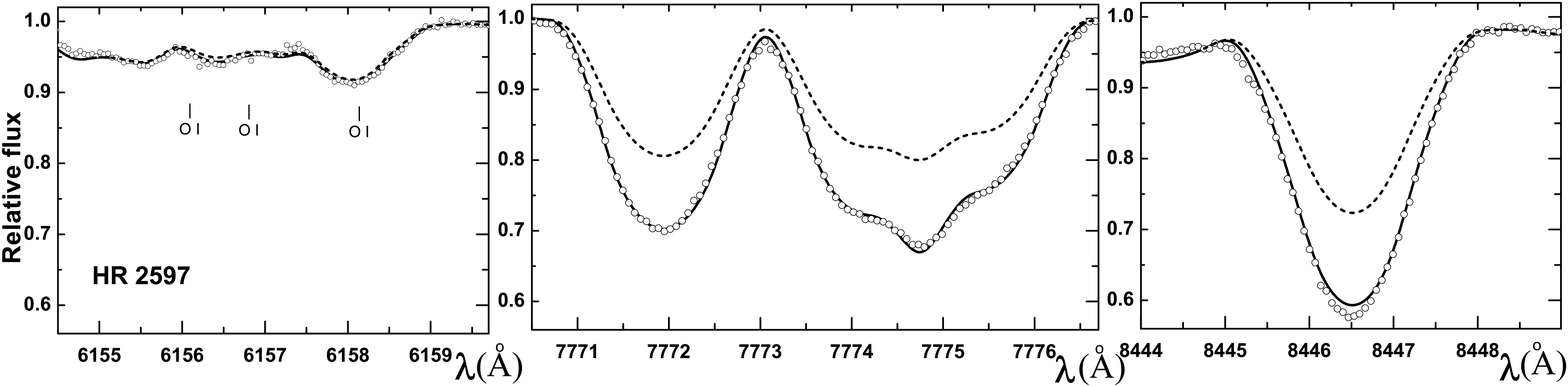}
\includegraphics[width=18cm,clip=true]{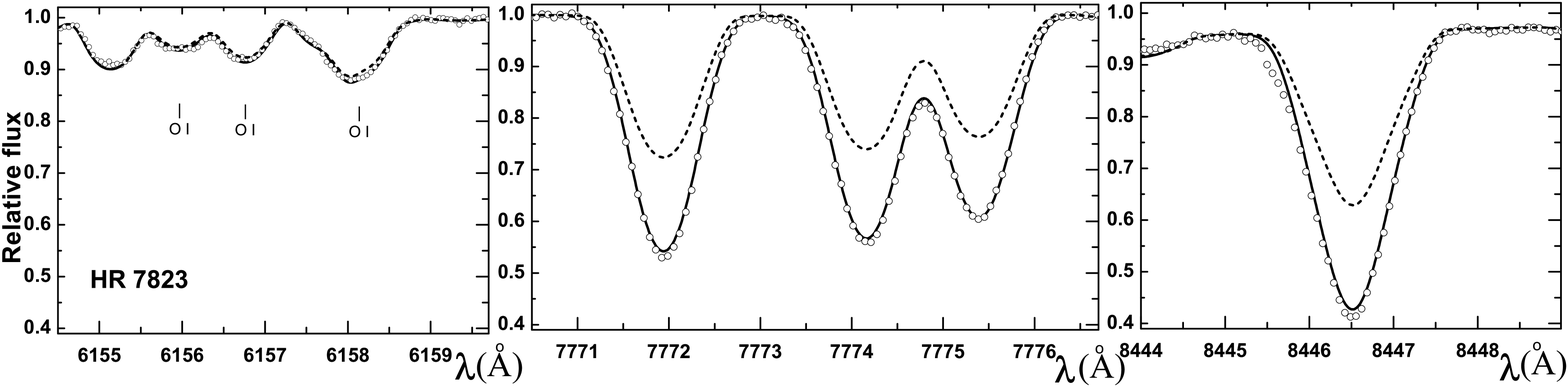}
\caption{Comparison of the observed and computed profiles of some \ion{O}{i}
lines for the stars HR~1327, 2597 and 7823. Solid curves - non-LTE, dotted 
curves - LTE. Computed profiles are obtained for the same abundance for each 
star (see Table \ref{star}). The dashed line is hidden under the solid line 
for the 5577 \AA\ and 6300 \AA, and mostly hidden in the case of the 
6155-6158 \AA, since they form close to LTE conditions.}
\label{prof}
\end{figure*}

Populations of \ion{O}{i} levels were determined using the MULTI code of 
\citet{Carlsson86} with modifications as given by \citet{Korotin99}. Proper 
comparison of observed and computed profiles 
in many cases requires a multi-element spectral synthesis in order to take into 
account possible blending lines of other species. For this process, we fold the 
non-LTE (MULTI) computations, specifically the departure coefficients, into the LTE
synthetic spectrum code SYNTHV \citep{Tsymbal96} that enables us to calculate the 
non-LTE source function and opacity for \ion{O}{i} lines. These calculations included all spectral 
lines from the VALD database \citep{Ryabchikova15} in a region of interest. 
The LTE approach was applied for lines other than the \ion{O}{i} lines. 
Abundances of corresponding chemical elements were adopted in accordance with 
the [Fe/H] values for each programme star. 

The oxygen model atom used in our non-LTE computations was first described by 
\citet{Mishenina00} and then updated by \citet{Korotin14}. The model 
consists of 51 \ion{O}{i} levels of singlet, triplet, and quintet systems, and the 
ground level of the \ion{O}{ii} ion. An additional 24 levels of neutral oxygen and 15 
levels of ions \ion{O}{ii} and \ion{O}{iii} are added for particle number conservation. Fine 
structure splitting was taken into account only for the ground level and the 
3p~5P level (the upper level of the 7771-7775~\AA\ triplet lines). A total of 248 
bound-bound transitions were included. Accurate quantum mechanical calculations 
were employed for the first 19 levels to find collision rates with electrons 
\citep{Barklem07}. 
The \ion{O}{i} model is described in detail by \citet{Korotin14}.

The IR triplets 7771-7775~\AA, 8446~\AA\ and 9260-9266~\AA\ show the greatest non-LTE 
effects. On the contrary, the forbidden lines 5577 and 6300~\AA\ are practically 
insensitive to non-LTE effects, so they can be considered as very suitable lines 
for an abundance analysis. However, there are some limitations on their application. For 
instance, the forbidden 6300~\AA\ line becomes too weak for stars with 
temperatures \teff $>$ 6500~K. Besides, for some programme stars with relatively 
high rotational velocities \vsini\ this line is strongly blended by the \ion{Sc}{ii}
6300.69 \AA\ line. Another forbidden line, 5577~\AA, has low intensity for all 
programme stars, so it was used as a subsidiary line in our analysis. 

The lines 6155, 6156 and 6158~\AA\, each of them is a triplet, showed a moderate 
departures from LTE. They were successfully used in the O abundance 
determination for a part of the stars. Unfortunately, for stars with 
\teff $<$ 5800~K these lines become too weak and, therefore, unusable for our 
analysis. 

We show in Fig. \ref{prof}  comparisons between observed and computed 
profiles of some \ion{O}{i} lines for three programme stars with the effective 
temperatures \teff = 5440, 6710 and 6760 K for HR~1327, 2597 and 7823, 
respectively. One may see that there is an excellent agreement in the non-LTE 
case (solid curves), whereas in the LTE case (dashed curves) for IR triplets 
the agreement is absent: the computed LTE profiles of the strong IR lines are much weaker than the non-LTE profiles fitted to the
observed profiles and which also fit the weak visual region permitted and forbidden lines where the LTE and non-LTE profiles are very similar. The case of HR~2597 is interesting because of the relatively 
high rotational velocity \vsini = 35 km/s of this star. Therefore, the strong IR 
lines for HR~2597 are significantly broadened in comparison with the star 
HR~7823 that has similar \teff\ but lower \vsini =19 km/s. 

One may see in Fig. \ref{prof} a number of interesting details. In particular, the IR 
lines are significantly strengthening when \teff\ increases. At the same time, 
when \teff\ increases the lines on 6155-6158~\AA\ become accessible for the 
analysis, whereas the forbidden lines at 6300 and 5577~\AA\ are weakened and 
become unuseable. 

\begin{figure}
%\includegraphics[width=\columnwidth,clip=true]{w_teff.eps}
%\caption{Equivalent widths of the lines 7771 \AA\ and 6158 \AA\ computed for 
\includegraphics[width=\columnwidth,clip=true]{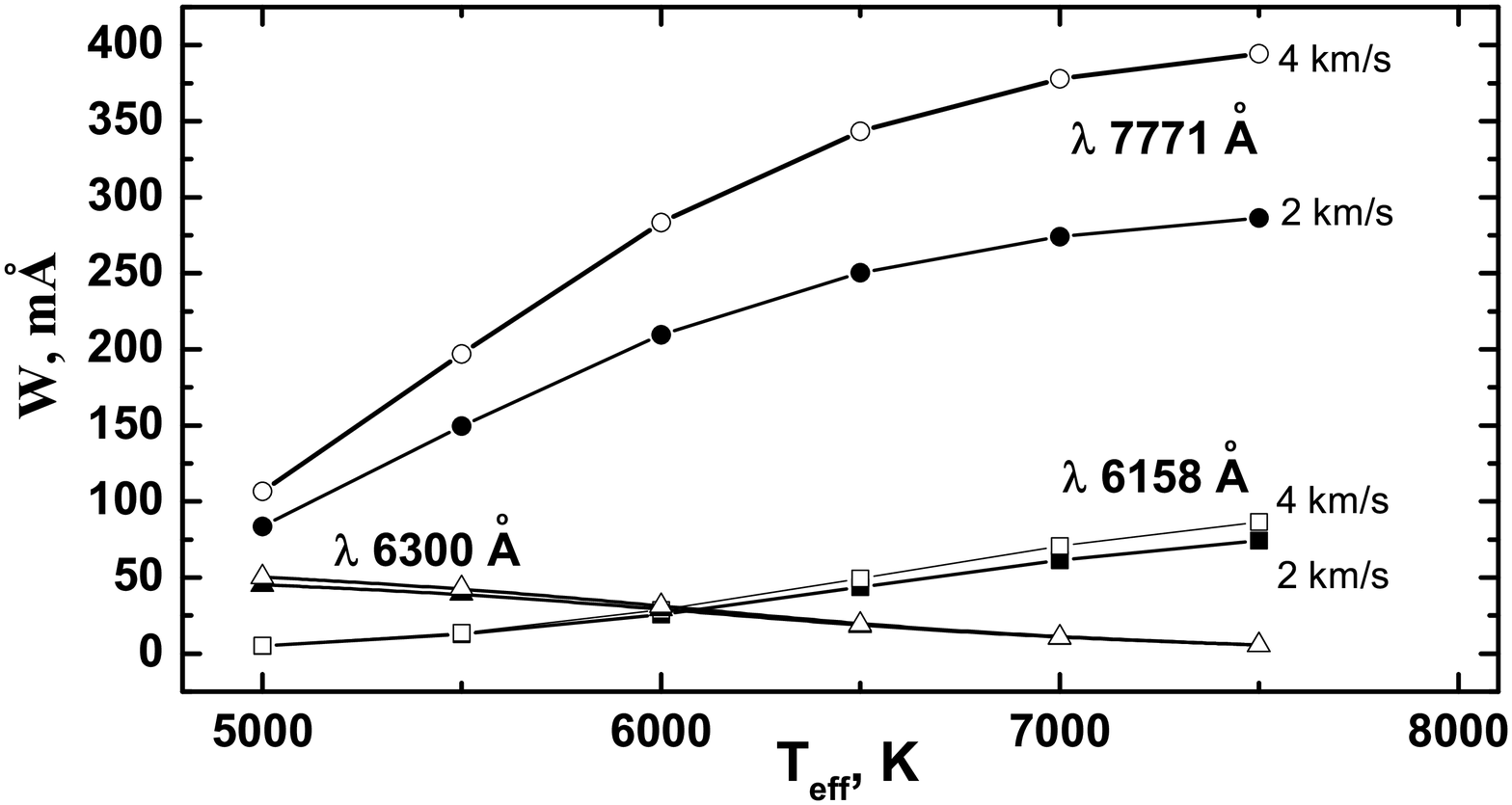}
\caption{Equivalent widths of the lines 7771~\AA, 6300~\AA\ and 6158~\AA\ computed for 
\vt = 2 and 4 km/s as a function of \teff.}
\label{wteff}
\end{figure}

\begin{figure}
\includegraphics[width=\columnwidth,clip=true]{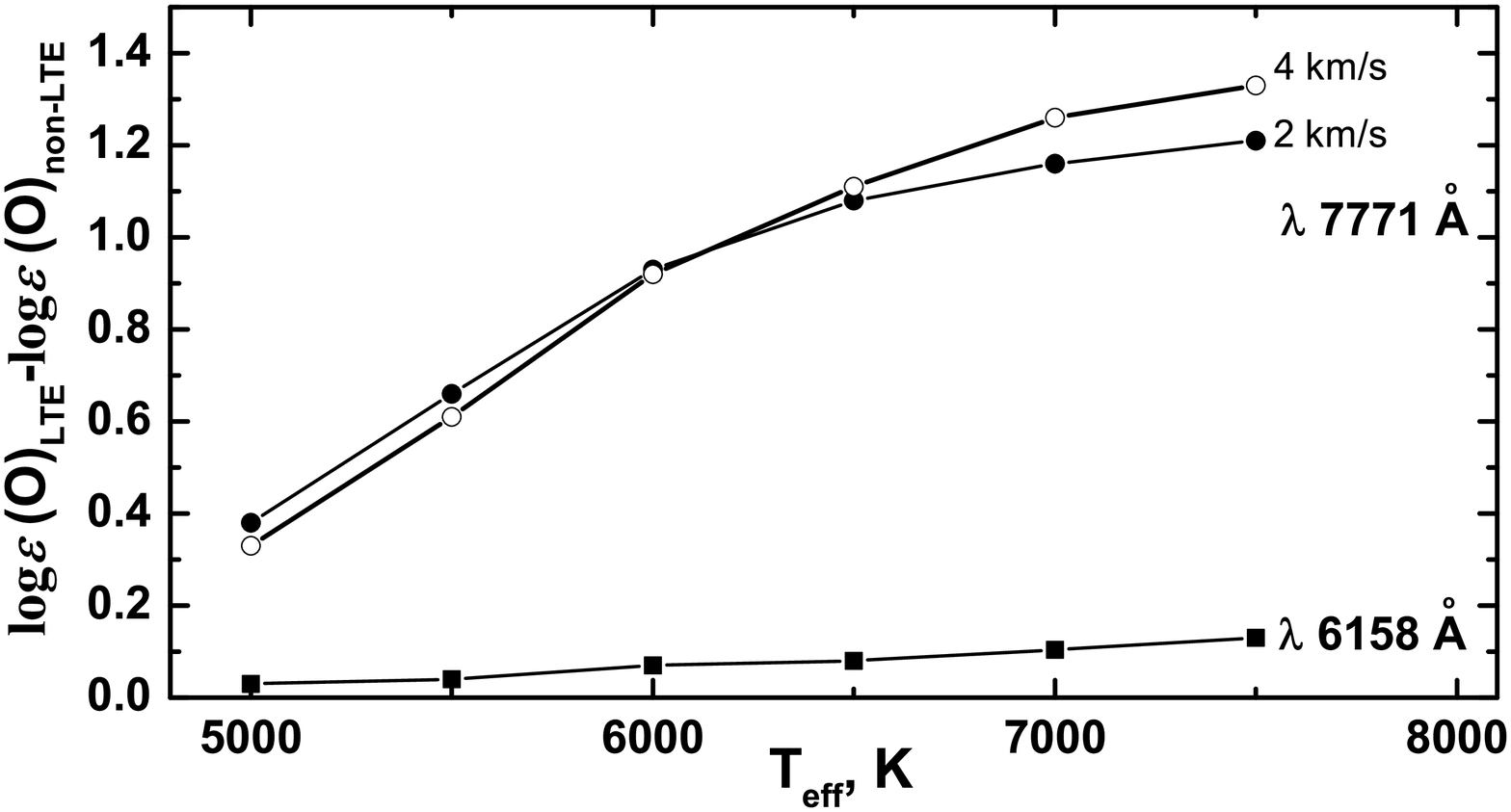}
\caption{The LTE--non-LTE difference in the oxygen abundance for the lines 7771 \AA\
and 6158 \AA\ as a function of \teff. Here \vt = 2 and 4 km/s for 7771 \AA, 
whereas for 6158 \AA\ the curves for \vt = 2 and 4 km/s are identical.}
\label{dnlteteff}
\end{figure}

We found that high-exitation IR triplets, on the one hand, and the 
low-exitation forbidden \ion{O}{i} lines in the visual 
region, on the other hand, show a different sensitivity to variations of 
star's parameters, especially to the \teff\ variations.
% and the microturbulence \vt\. 
We illustrate this  with the help of Figs. \ref{wteff} and \ref{dnlteteff}, 
where some results of our computations for lines 7771~\AA, 6300~\AA\ and 
6158~\AA\ are presented. In particular, Fig. \ref{wteff} shows equivalent 
widths W of these lines as a function of \teff, whereas Fig. \ref{dnlteteff} 
shows a difference between LTE and non-LTE abundances as a function of 
\teff\ (non-LTE corrections for 6300 \AA\ are zero, so they are not 
shown in Fig. \ref{dnlteteff}).
Note that the computations were implemented for \logg = 2.0, a solar O 
abundance and two \vt\ values. One may see from Figs. \ref{wteff} and 
\ref{dnlteteff} that both the equivalent width W and the non-LTE correction in 
O abundances for the IR line 7771~\AA\ are significantly strengthened with 
\teff, whereas for the line 6158~\AA\ the strengthening is small.
On the contrary, the $W$ values are weakened for the low-exitation forbidden 
6300 \AA\ line.

It is interesting to note that the non-LTE computations of \ion{O}{i} lines for FGK 
dwarfs were made recently by \citet{Amarsi16} on the basis of 3D 
hydrodynamic model stellar atmospheres. Unfortunately, we cannot compare 
directly the results of our non-LTE computations based on 1D hydrostatic models with these 
data because our programme stars (supergiants and bright giants) have lower \logg\ 
values.

\section{DETERMINATION OF THE OXYGEN ABUNDANCES}

We applied in this work, as in Papers I-IV, the model atmospheres computed by us 
for the adopted \teff, \logg\ and \vt\ values on the basis of \citet{Kurucz93} 
ATLAS9 code. Using these model atmospheres and the microturbulent parameter \vt\ 
from Table \ref{star}, we determined the oxygen abundance log $\varepsilon$(O) for each star. The 
log $\varepsilon$(O) values are given in the standard scale, where for hydrogen the value 
log $\varepsilon$(H) = 12.00 is adopted.  As noted above, in order to take into account the 
blending by lines of other chemical elements, we based our analysis on 
computations of synthetic spectra and their comparison with observed ones.
For instance, it is known that in the solar spectrum the 5577 \AA\ line is 
severely blended with C$_{2}$ lines \citep{Melendez08}, 
the 6300 \AA\ line is severely blended with \ion{Ni}{i} \citep{Prieto01}, 
and the 8446 \AA\ lines are affected by two \ion{Fe}{i} lines. 
Effects of these lines took into consideration in our computations.

Computation of synthetic spectra requires an adoption of the projected rotational 
velocity \vsini\ for each star. It was noted in Paper III that when spectra of 
cool supergiants and bright giants are studied, it is difficult to separate 
correctly the contributions to a line profile of the projected rotational 
velocity \vsini\ and the macroturbulent velocity $V_{\text {mac}}$. Two limiting cases  
were considered in Paper III: calculation of spectra with only \vsini\ or only 
with $V_{\text {mac}}$; it was found that the derived Li abundances for these two cases are 
identical within 0.03 dex. Noting this result, we used here spectra computed using the \vsini\ values and neglected $V_{\text {mac}}$. The \vsini\ values were 
determined in Papers III and IV from Li I and C I lines, respectively, with  
 excellent agreement between the two determinations. We displayed already in 
Fig. \ref{prof}, as an example of our technique, the fitting of the computed non-LTE 
profiles of some \ion{O}{i} lines to the observed ones for three stars.  

% **********************
We implemented a preliminary non-LTE analysis of the \ion{O}{i} lines from 
Table \ref{line} for all 63 stars listed in Paper I. As a result of such 
consideration, some stars were excluded from further analysis (see below). We 
included in a final list only the stars, for which a good agreement is obtained 
between the abundances log $\varepsilon$(O) determined for three groups of the 
\ion{O}{i} lines: 
1) the low-excitation forbidden lines on 6300 \AA\ and 5577 \AA, which are weak 
and blended, but insensitive to departures from LTE; 
2) the high-excitation IR triplets 7771-7775 \AA\ and 8446 \AA, which are strong 
and less sensitive to blends, but can show large departures from LTE; 
3) the high-excitation lines on 6155-6158 \AA, which show weaker departures 
from LTE. We selected 51 stars in all; their oxygen abundances 
log $\varepsilon$(O) are presented in the last column of Table \ref{star}. It 
should be noted that all these stars have distances d $\leq$ 1000 pc; 
only for HR 7876 d $\approx$ 1700 pc (see Paper I). 

The omitted 12 stars showed a marked discrepancy in the log $\varepsilon$(O) 
values (about 0.3 dex or more) between the \ion{O}{i} lines in IR and visual 
regions; some discrepancy takes place as well between the forbidden and 
permitted lines in the visual region. In particular, such situation was 
obtained for two relatively hot A5-type supergiants, HR~825 (Teff = 8570 K) 
and HR~2874 (Teff = 8620 K), for which departures from LTE in IR lines are 
especially great. Apart from the possible uncertainties in the \teff\ values for 
these supergiants, errors in their high microturbulent parameters, \vt = 10.8 
and 7.8 km/s, can distort the results. Note as well that a possible dependence 
of \vt\ on height in atmospheres of such luminous stars can play a marked role.
Such dependence was found, for example, for F supergiants by \citet{Lyubimkov90}. 

There are also several cooler stars, for which we cannot find the O abundance 
with confidence. A marked discrepancy between \ion{O}{i} lines in IR and visual 
regions was obtained for them, too. Possible errors in the parameters \teff\ and 
\vt\ can be a reason of the discrepancy. It should be noted that our results for 
oxygen on the whole seem to be more sensitive to the Teff and Vt variations then 
the previous results for carbon and nitrogen. We omitted 12 stars with 
unreliable log $\varepsilon$(O) in Table \ref{star}; however, we hope to 
redetermine in future more exactly both their basic parameters and the CNO 
abundances.  
% *****************************

When deriving the oxygen abundance log $\varepsilon$(O) for 51 stars in 
Table \ref{star}, we used from 4 to 10 lines for each star. We evaluated that 
the mean errors in log $\varepsilon$(O) from the analyzed \ion{O}{i} lines vary
from 0.07 to 0.22 dex, if the uncertainties $\Delta$ \teff = $\pm$ 100 K, 
$\Delta$ \logg = $\pm$ 0.2 and $\Delta$ \vt = $\pm$ 0.3 km/s are adopted. The 
errors $\sigma$ in log $\varepsilon$(O) presented in Table \ref{star} take into 
account as well the uncertainty in fitting of computed to the observed profiles. 

\begin{figure}
\includegraphics[width=\columnwidth,clip=true]{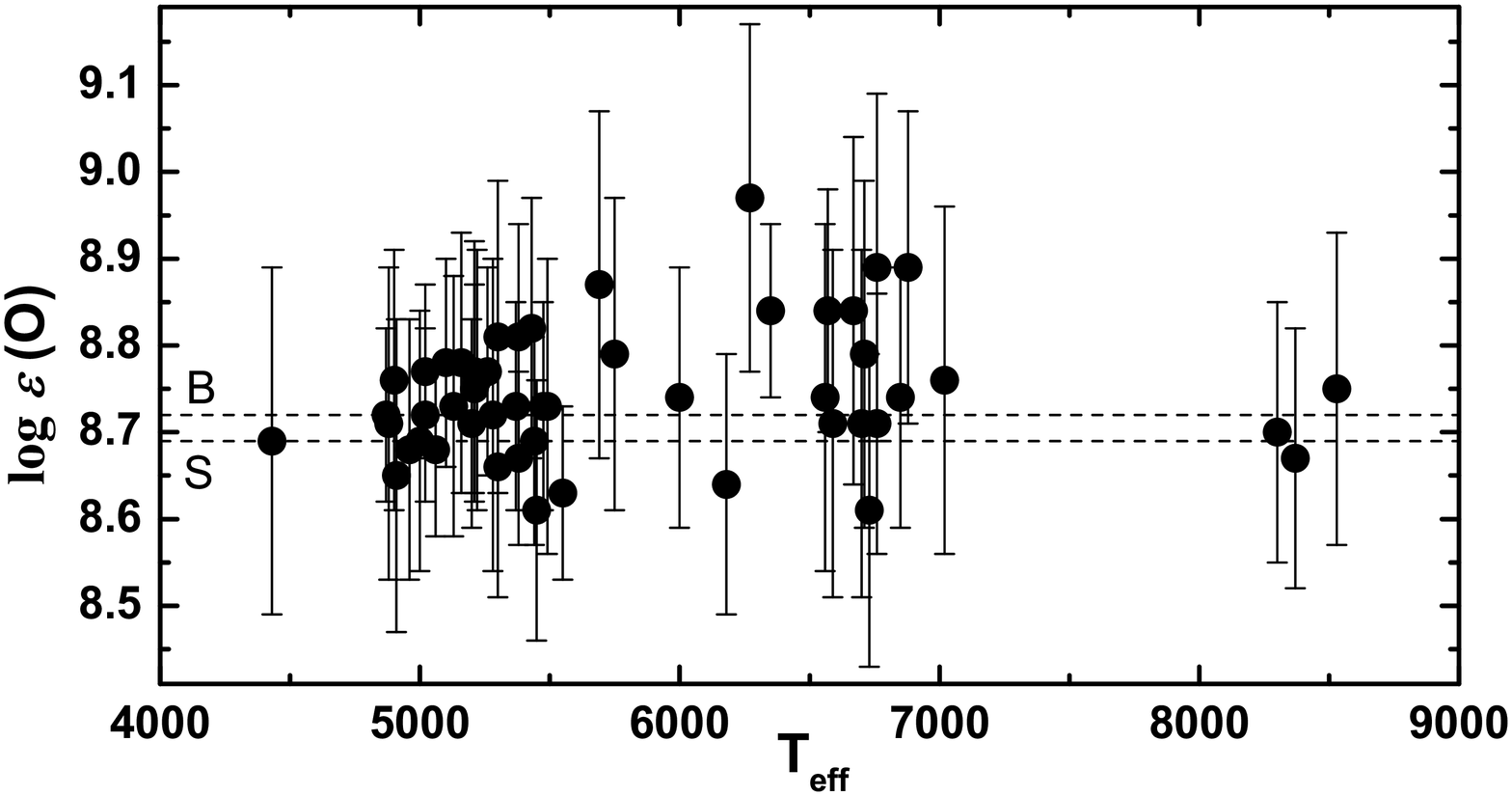}
\includegraphics[width=\columnwidth,clip=true]{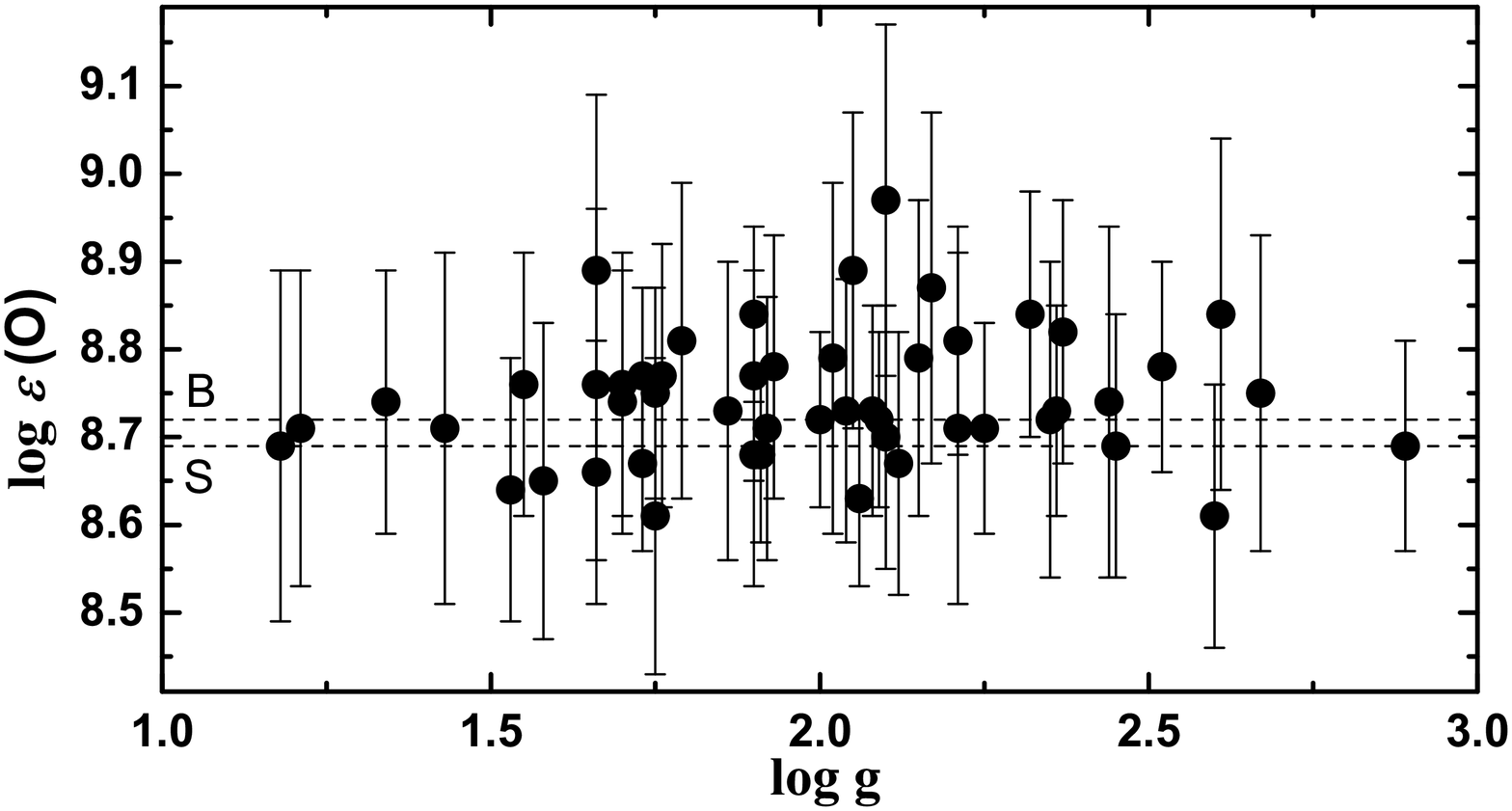}
\caption{The derived oxygen abundance for 51 programme stars as a function of 
\teff\ and \logg. Two important log $\varepsilon$(O) values are shown also by 
dashed lines, namely S - the solar O abundance log $\varepsilon$(O) = 8.69 
\citep{Asplund09}; B - the mean O abundance log $\varepsilon$(O) = 8.72 for 
the unevolved early B-type MS stars \citep{Lyubimkov13}.}
\label{o_ab}
\end{figure}

In Fig. \ref{o_ab} we show the derived oxygen abundance log $\varepsilon$(O) for programme stars as 
a function of two basic stellar parameters, namely the effective temperature 
\teff\ (upper panel) and the surface gravity \logg\ (lower panel). Two important 
values are shown by dashed lines: first, S - the solar O 
abundance log $\varepsilon$(O) = 8.69 \citep{Asplund09}; second, B - the mean O 
abundance log $\varepsilon$(O) = 8.72 for the unevolved early B-type MS stars 
\citep{Lyubimkov13}. As known, the CNO-abundances for such stars can be 
considered as the initial CNO-abundances for AFG supergiants and bright giants, 
descendants of  B-type MS stars.

One may see from Fig. \ref{o_ab} that the derived log $\varepsilon$(O) values 
agree within their error bars  with the S and B lines that correspond to the initial O abundance. 
Moreover, there is no trend with \teff\ or \logg\ across the \teff\ interval from 
4500 to 8500~K and the \logg\ interval from 1.2 to 2.9 dex. 
We may conclude that, within possible uncertainties, the O abundance in the 
stars studied did not change markedly on either the main sequence  or in 
evolution to the AFG supergiant stage. 

Lack of a tie between O abundance and a star's evolutionary stage contrasts 
sharply with results for the C and N abundances found in Papers II and IV, 
i.e., carbon deficiencies of between $-0.1$ and $-0.7$ dex and nitrogen 
overabundances of $+0.2$ to $+0.9$ dex in the supergiants, as compared with 
the Sun,  and crucially a tight correlation between the C deficiency and the N 
overabundance. The contrast between C-N correlation and the independence of O 
on the evolutionary stage is readily understood in qualitative terms. First, 
the CN-cycle component of the H-burning CNO-cycles operates much faster than 
the ON-cycle. Thus, the CN-cycle converts C to N while leaving O almost 
completely unaffected. Operation of the ON-cycle, generally requiring  higher 
temperatures, converts O to N but may raise the N abundance considerably above 
that provided by the CN-cycle; the initial O abundance is a factor of two 
greater than the initial C abundance. Second, predictions about changes to
surface abundances of C, N and O caused by rotationally-induced mixing confirm 
the  weaker role for the ON-cycle relative to the CN-cycle. For example, the 
computations of \citet{Heger00} predict that for AFG supergiants in the 
post-FDU phase the log $\varepsilon$(O) may be lowered by $<$ 0.05 dex 
for \v0 = 200 km/s and $<$ 0.15 dex for \v0 = 300 km/s. Only for rare stars 
with high rotational velocities \v0 $\sim$ 400 km/s the log $\varepsilon$(O) 
changes become observable. Predicted changes to C and N abundances are much 
larger, anti-correlated and comparable to the observed values.

\section{THE N/C vs N/O RELATION}

In pursuit of a quantitative comparison between observations and predictions of the mixing to their surface of products of the H-burning CNO-cycles -- both the CN-cycle and the ON-cycle -- we utilize  the N/C vs N/O relation which may be considered  a  sensitive 
indicator of products of the CNO-cycles \citep{Maeder14}. In order to construct this relation,  it 
is necessary to have abundances of all three elements - C, N and O. 
Along with abundances obtained for oxygen in the present work, we used the previous 
data for nitrogen and carbon from Papers II and IV.
As a result, we have the C, N and O abundances for 17 stars, i.e., for 1/3 of 
51 stars analyzed for O in the present work. 

These 17 stars and their C, N and O abundances are presented in 
Table \ref{CNO}. The abundances are given relatively to solar abundances. For example, 
the [C/Fe] value is defined as [C/Fe] = log (C/Fe)$_{\text {star}}$ -- log (C/Fe)$_{\text {sun}}$, 
where log (C/Fe) = log $\varepsilon$(C) -- log $\varepsilon$(Fe). When 
calculating [C/Fe], [N/Fe] and [O/Fe] for stars in Table \ref{CNO}, we adopted the 
solar abundances C, N, O and Fe from \citet{Asplund09} review. 
It should be noted that thanks to  use of relative values [C/Fe], [N/Fe] and 
[O/Fe] instead of the standard abundances log $\varepsilon$(C), 
log $\varepsilon$(N) and log $\varepsilon$(O) we may compensate for differences 
in the metallicity [Fe/H] between stars.

\begin{table}
%\centering
\caption{The C, N and O abundances and the [N/C] and [N/O] values for 17 stars.}
\label{CNO}
\begin{tabular}{lcccccc}
\hline
 HR &[C/Fe]&[N/Fe]&[O/Fe]&[N/C]  & [N/O]& log $\varepsilon$(C+N+O)\\
\hline
\hline
 292&-0.34&0.61& 0.13&0.95$\pm$0.17& 0.48$\pm$0.22&9.09$\pm$0.25\\
1017&-0.12&0.65& 0.20&0.77$\pm$0.16& 0.45$\pm$0.14&9.05$\pm$0.19\\
1740&-0.12&0.37& 0.07&0.49$\pm$0.13& 0.30$\pm$0.17&8.90$\pm$0.20\\
1865&-0.69&0.87& 0.00&1.56$\pm$0.17& 0.87$\pm$0.17&9.06$\pm$0.23\\
2597&-0.21&0.50& 0.26&0.71$\pm$0.38& 0.24$\pm$0.35&8.94$\pm$0.43\\
3102&-0.24&0.41& 0.05&0.65$\pm$0.32& 0.36$\pm$0.32&9.07$\pm$0.38\\
3183&-0.08&0.25& 0.00&0.33$\pm$0.16& 0.25$\pm$0.18&8.97$\pm$0.24\\
6081&-0.03&0.44&-0.07&0.47$\pm$0.09& 0.51$\pm$0.17&8.97$\pm$0.18\\
6978&-0.30&0.51& 0.12&0.81$\pm$0.13& 0.39$\pm$0.17&8.92$\pm$0.20\\
7014&-0.13&0.60& 0.25&0.73$\pm$0.22& 0.35$\pm$0.26&9.07$\pm$0.30\\
7094&-0.16&0.51& 0.09&0.67$\pm$0.22& 0.42$\pm$0.24&8.83$\pm$0.28\\
7264&-0.18&0.64& 0.17&0.82$\pm$0.23& 0.47$\pm$0.23&8.92$\pm$0.31\\
7387&-0.40&0.81& 0.03&1.21$\pm$0.20& 0.78$\pm$0.24&9.01$\pm$0.28\\
7770&-0.11&0.53& 0.15&0.64$\pm$0.19& 0.38$\pm$0.19&8.85$\pm$0.24\\
7823&-0.23&0.73& 0.13&0.96$\pm$0.21& 0.60$\pm$0.21&8.95$\pm$0.25\\
7834&-0.21&0.37& 0.13&0.58$\pm$0.19& 0.24$\pm$0.18&9.01$\pm$0.23\\
7876& 0.00&0.13& 0.26&0.13$\pm$0.18&-0.13$\pm$0.24&8.90$\pm$0.27\\
\hline                                                         
\end{tabular}                                                  
\end{table}                                                   

\begin{figure}
\includegraphics[width=\columnwidth,clip=true]{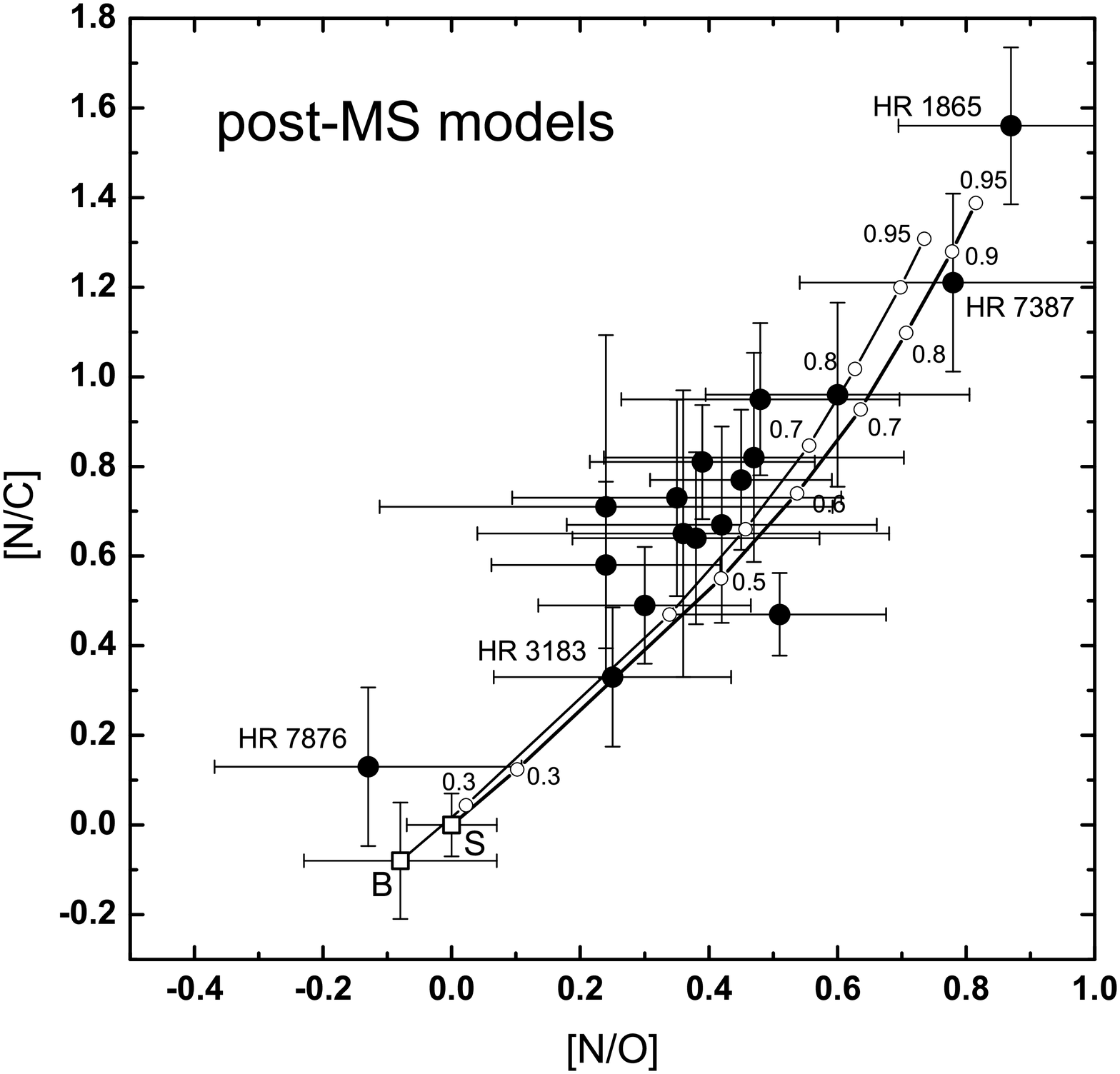}
\includegraphics[width=\columnwidth,clip=true]{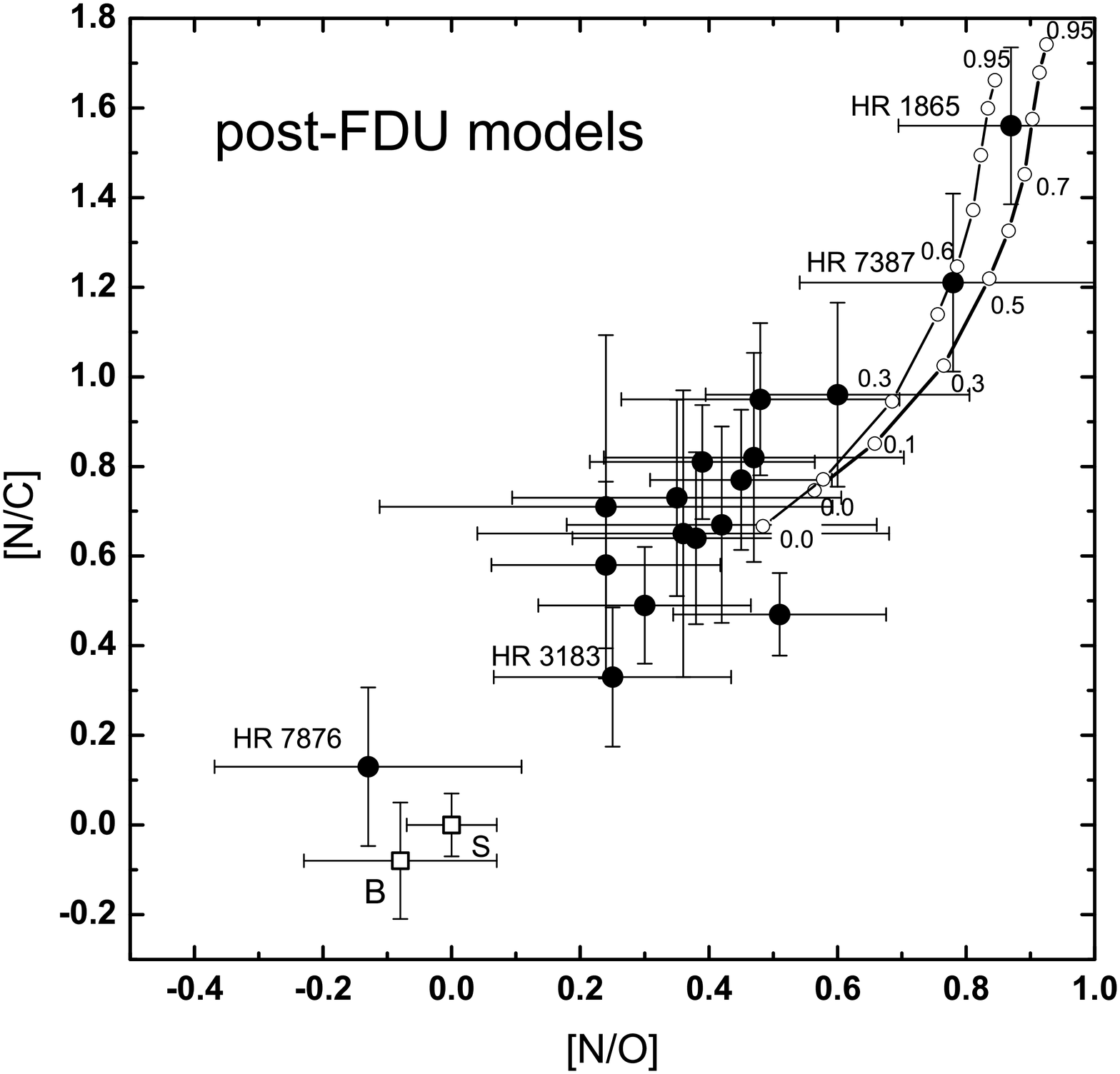}
\caption{Comparison of the observed [N/C] vs [N/O] relation with theoretical 
predictions of \citet{Georgy13} for the 9 M$_{\sun}$ model. Open squares S and 
B correspond to the same log $\varepsilon$(O) values for the Sun and B stars as
in Fig. \ref{o_ab}.  Solid lines correspond to predictions for the post-MS phase (upper panel) and the post-FDU phase (lower panel). Initial relative angular velocities 
$\Omega$/$\Omega_{cr}$ are marked near by the corresponding nodal points. 
Positions of the stars HR~1865, 3183, 7387 and 7876 are marked.}
\label{msfdu}
\end{figure}

The [N/C] and [N/O] values in Table \ref{CNO} were defined according to the following 
formulae: [N/C] = [N/Fe] -- [C/Fe] and [N/O] = [N/Fe] -- [O/Fe]. The derived 
[N/C] and [N/O] values (filled circles) and their error bars are shown in 
Fig. \ref{msfdu}. One may see that there is an obvious correlation between [N/C] and [N/O]. 
The [N/C] increase from 0 to 1.6 dex is accompanied by the [N/O] increase from 
0 to 0.9 dex. 

It is interesting to compare this observed relation with predictions of the 
theory. We used for such comparison computation of \citet{Georgy13} for 
rotating stellar models. Results for two evolutionary phases, post-MS and 
post-FDU, are shown in Fig. \ref{msfdu} separately (upper and lower panels, respectively). 
It should be noted that Georgy et al. obtained their C, N and O abundances 
as mass fractions; we converted their values into our scale (by number of atoms). 
Note as well that we used their results obtained for the solar metallicity $Z$ = 0.014 
(see their Table 2); we identified their ''End of H-burning`` as the post-MS phase and 
their ''End of He-burning`` as the post-FDU phase.

\citet{Maeder14} noted that the theoretical [N/C] vs [N/O] relation is 
slightly dependent on the stellar mass M. Therefore, we show in Fig. \ref{msfdu} the 
theoretical data for the mass $M$ = 9 M$_{\sun}$, that is approximately the 
mean mass for stars of our sample. Nodal points on the theoretical tracks 
(open circles) in Fig. \ref{msfdu} correspond to the various initial angular velocities 
$\Omega$/$\Omega_{cr}$ from 0 to 0.95, where $\Omega_{cr}$ is a critical 
angular velocity. The relation between the $\Omega$/$\Omega_{cr}$ 
values and the corresponding equatorial velocities \v0 is presented in Table \ref{rot}. 
One may see from Fig. \ref{msfdu} that theoretical [N/C] and [N/O] values depend strongly 
on the initial rotational velocity.

\begin{table}
\centering
\caption{Initial rotational velocities in the 9 M$_{\sun}$ model. 
Here $\Omega$/$\Omega_{cr}$ - relative angular velocity, 
\v0\ - corresponding equatorial velocity (\citealt{Georgy13}, 
see their Table 2.)}
\label{rot}
\begin{tabular}{cc}
\hline
$\Omega$/$\Omega_{cr}$&	\v0 (km/s)\\                                    
\hline
\hline
0.00&0   \\
0.10&40  \\
0.30&130 \\
0.50&216 \\
0.60&261 \\
0.70&311 \\
0.80&381 \\
0.90&466 \\
0.95&492 \\
\hline                                                         
\end{tabular}                                                  
\end{table}                                                   
    
In order to consider the N/C vs N/O relation as an evolutionary phenomenon, we 
should indicate the initial values log(N/C) = log $\varepsilon$(N) -- log $\varepsilon$(C) and 
log (N/O) = log $\varepsilon$(N) -- log $\varepsilon$(O). We adopted at first Asplund et al.'s (2009) 
data for the Sun, namely log(N/C) = --0.60 and log (N/O) = --0.86; remind that 
the [N/C] and [N/O] values in Table \ref{CNO} were related just to these 
Asplund et al.'s values. It is important that in Georgy et al.'s computations 
presented in Fig. \ref{msfdu} practically the same initial values log(N/C) and log(N/O) 
have been adopted. In Fig. \ref{msfdu} the point S with zero coordinates corresponds to 
these initial values. 

It should be noted that the solar C, N and O abundances are known presently 
with high accuracy. For instance, if one takes \citet{Caffau11} data 
instead of Asplund et al.'s values, the corresponding ''zero point``  in Fig. \ref{msfdu} 
would have the coordinates [N/C] = --0.04 and [N/O] = --0.04. Next, if we adopt 
the C, N and O abundances determined by us for the Sun with our non-LTE 
technique, ''zero point`` would have the coordinates [N/C] = --0.06 and 
[N/O] = --0.06. 

Apart from these very close solar data, we used the above-mentioned 
CNO-abundances determined for the unevolved early B-type MS stars 
\citep{Lyubimkov13}. These data can be actually considered as the initial 
abundances for AFG supergiants and bright giants, which are descendants of the 
B-type MS stars. The corresponding ''zero point`` B in Fig. \ref{msfdu} has the coordinates 
[N/C] = --0.08 and [N/O] = --0.08. Note that ''zero points`` S and B correspond to 
the zero initial velocity \v0 = 0 km/s. 

One may see that two theoretical curves with beginning at the points S and B 
are very close. On the whole, from the qualitative point of view, there is a 
good agreement between these theoretical predictions and the observed 
[N/C] vs [N/O] relation. Both theoretical curves lie within error bars 
in Fig. \ref{msfdu}. Positions of the curves show a little shift and remain within 
error bars, if we use \citet{Georgy13} computations for other masses $M$ 
instead of $M$ = 9 M$_{\sun}$. We arrive at the same conclusion, when using earlier 
computations, for instance, \citet{Heger00} ones. 

It is obvious that this relation is explained by the 
strong dependence on the initial rotational velocity \v0: the higher the 
velocity \v0\ the greater on average the values [N/C] and [N/O]. Note that the 
same explanation, i.e. dependence on \v0, has been found in Paper IV for the
N vs C anticorrelation in AFG supergiants and bright giants. These phenomena 
reflect the presence on a surface of rotating stars of the mixed material from 
stellar interiors. So, a majority of these stars passed through the 
rotationally induced mixing during the MS stage and/or the FDU phase. 

As far as the quantitative agreement between theory and observation in Fig. \ref{msfdu}, 
a detailed discussion is needed. We shall discuss as well our earlier relations for 
CNO abundances and an accordance of the relations obtained with the expected 
initial rotational velocities of stars.

\section{DISCUSSION}

A role of rotation seems to be very important. In this connection the following
question arises: what the real initial rotational velocities \v0\ occur for stars
in question? In other words, what is it known about the \v0\ values for the early 
B-type MS stars that are progenitors of AFG supergiants? We note once again that 
the stars of both types have masses $M$ = 4-19 M$_{\sun}$. The problem of 
observed rotational velocities for the MS stars of various spectral types is 
discussed in \citet{Lyubimkov18} review on light elements in stars. 

In particular, it is noted there that modern observational data show that most 
of early B stars (about 80 \%) had low rotational velocities 
$\sim$ 0-150 km/s at the beginning of evolution on MS. Moreover, a 
substantial fraction of them fell within a still narrower interval, 0-50 km/s. 
On the other hand, the number of stars with relatively high rotational 
velocities from 150 to 300-400 km/s is small (about 20 \%). These results are 
very important for an interpretation of the observed evolutionary changes in 
abundances of light elements including the C, N and O abundances. 

As an example of good agreement with these results, one may indicate an 
analysis of the N/O ratio for 46 early B-type MS stars \citep{Lyubimkov16, Lyubimkov18}. 
It was found that most of the stars near the MS termination (82 \%) show the 
relatively low values [N/O] $\lid$ 0.3; from the viewpoint of the theory 
\citep{Georgy13}, these stars had the low initial rotational velocities 
\v0\ from 0 to 130 km/s. Only for 4 stars (18 \%) the enhanced values 
[N/O] = 0.4-0.8 were derived; the models with \v0 = 220-300 km/s correspond to 
them. So, this study displays a very good accordance between three independent 
sets of data, namely 1) the abundances determined from observed spectra; 2) the 
abundances computed from rotational models; 3) the expected initial rotational 
velocities for stars in question. 

Is there such accordance in the case of AFG supergiants and bright giants? In 
order to answer this question, it would be useful to remember the 
above-mentioned N vs C anticorrelation in these stars obtained by 
\citet{Lyubimkov15} (see Fig.12 there). Like the present paper, a comparison 
between the observed and computed relations has been implemented there for the 
same two phases: post-MS and post-FDU. It should be noted that the earlier 
computations of \citet{Heger00} were used there, because the calculated 
N and C abundances are not presented in \citet{Georgy13} work. It was 
found by \citet{Lyubimkov15} that the most of the stars forms on the 
observed N vs C relation a compact cluster with [C/Fe] between --0.1 and --0.4 
and with [N/Fe] between 0.3 and 0.7; they may be identified as either post-MS 
objects with \v0 $\approx$ 200-250 km/s or post-FDU objects with 
\v0 $\approx$ 0-150 km/s. So, there was some ambiguity in evolutionary status 
of these stars. 

Recently this problem was solved by \citet{Lyubimkov18} on the basis of the 
above-mentioned data on the initial rotational velocities of the stars. Since 
about 80 \% of the stars studied should have \v0 $<$ 150 km/s at the beginning 
of their evolution, it is obvious that the overwhelming majority of stars in 
the indicated cluster on the N vs C relation are the post-FDU objects with 
\v0 $\approx$ 0-150 km/s. The number of these stars is 19, i.e. 83 \% of a 
total number of the most probable post-FDU stars (23 in all). The remaining 4 
stars (17 \%) with the highest nitrogen abundances [N/Fe] = 0.8-0.9, according 
to rotating models, correspond to velocities \v0 $\approx$ 200- 300 km/s. One 
may conclude that the N vs C anticorrelation observed for AFG supergiants and 
bright giants also shows a good agreement both with the predicted N vs C 
relation and with the expected rotational velocities. 

Returning to Fig. \ref{msfdu}, we should remind that the initial relative angular 
velocities $\Omega$/$\Omega_{cr}$ are marked there nearby the nodal points on 
theoretical curves. These values vary from 0 (for ''zero points`` S and B) to 
0.95; the corresponding equatorial velocities \v0\ for the 9 M$_{\sun}$ model 
are presented in Table \ref{rot}. Basing on these \v0\ values, we may analyze 
consecutively both cases in Fig. \ref{msfdu}, i.e. both the post-MS phase  and the 
post-FDU phase. 

Basing on the above-mentioned data concerning the most probable initial 
rotational velocities \v0 $<$ 150 km/s for stars in question, we may see that 
only one star, HR~7876 (marked in Fig. \ref{msfdu}), corresponds exactly to this upper 
limit. It was shown in Paper IV that this star with the almost unchanged 
C and N abundances is the post-MS object with \v0 $\sim$ 100 km/s or less. 
One more star nearby HR~7876 in Fig. \ref{msfdu}, namely HR~3183 is close to the limit 
value \v0 = 150 km/s. However, majority of the remaining stars, 15 of 17, are 
located markedly higher than \v0 = 150 km/s. Only a little part of them can be 
in the post-MS phase; we may suppose that they are mostly in the post-FDU phase. 

Two F supergiants, HR~7387 and HR~1865 (marked in Fig. \ref{msfdu}), with the highest 
[N/C] values (1.21 and 1.56 respectively) are of special interest. In the case 
of HR~7387 a good agreement with model computations is obtained for 
\v0 $\approx$ 200-250 km/s, whereas in the case of HR~1865 ($\alpha$ Lep) the 
agreement is attained for \v0 $\sim$ 400 km/s. (Note that HR~1865 is the 
most massive star in Fig. \ref{msfdu}; its mass is 13.9 M$_{\sun}$). 

Locations of all other stars (13 in all), if they are post-FDU objects, 
correspond to the initial velocities \v0\ between 0 and 130 km/s. Nevertheless, 
it is necessary to note that among these 13 stars a little part 
(maybe 1-2 stars) can be the post-MS objects with \v0 $\approx$ 200- 300 km/s.
Then a total number of stars with \v0 $<$ 150 km/s in Fig. \ref{msfdu} would be about 80 \%
that is in excellent agreement with the observed distribution of stars on \v0. 	

Therefore, the observed C, N and O abundances in AFG supergiants and bright 
giants, in particular, the observed N vs C and N/C vs N/O relations show good 
agreement with predictions of modern theory, i.e. with computations of rotating 
models. 

An important test of reliability of the derived C, N and O abundances could be 
a conservation of the total C+N+O abundance during stellar evolution. In other 
words, according to the theory, the total value log $\varepsilon$(C+N+O) 
should remain the same from the beginning of the MS stage to the end of the 
AFG supergiant stage.

\begin{figure}
\includegraphics[width=\columnwidth,clip=true]{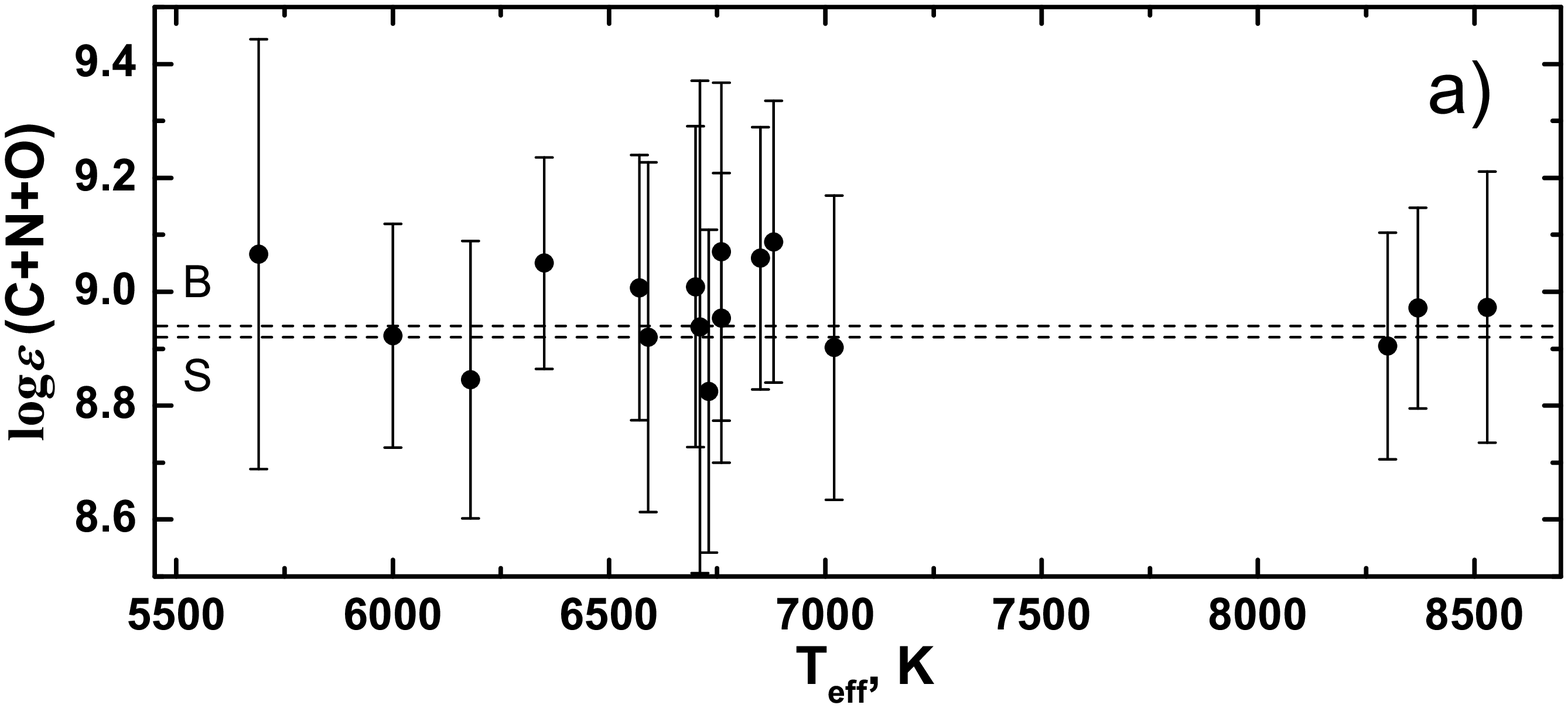}
\includegraphics[width=\columnwidth,clip=true]{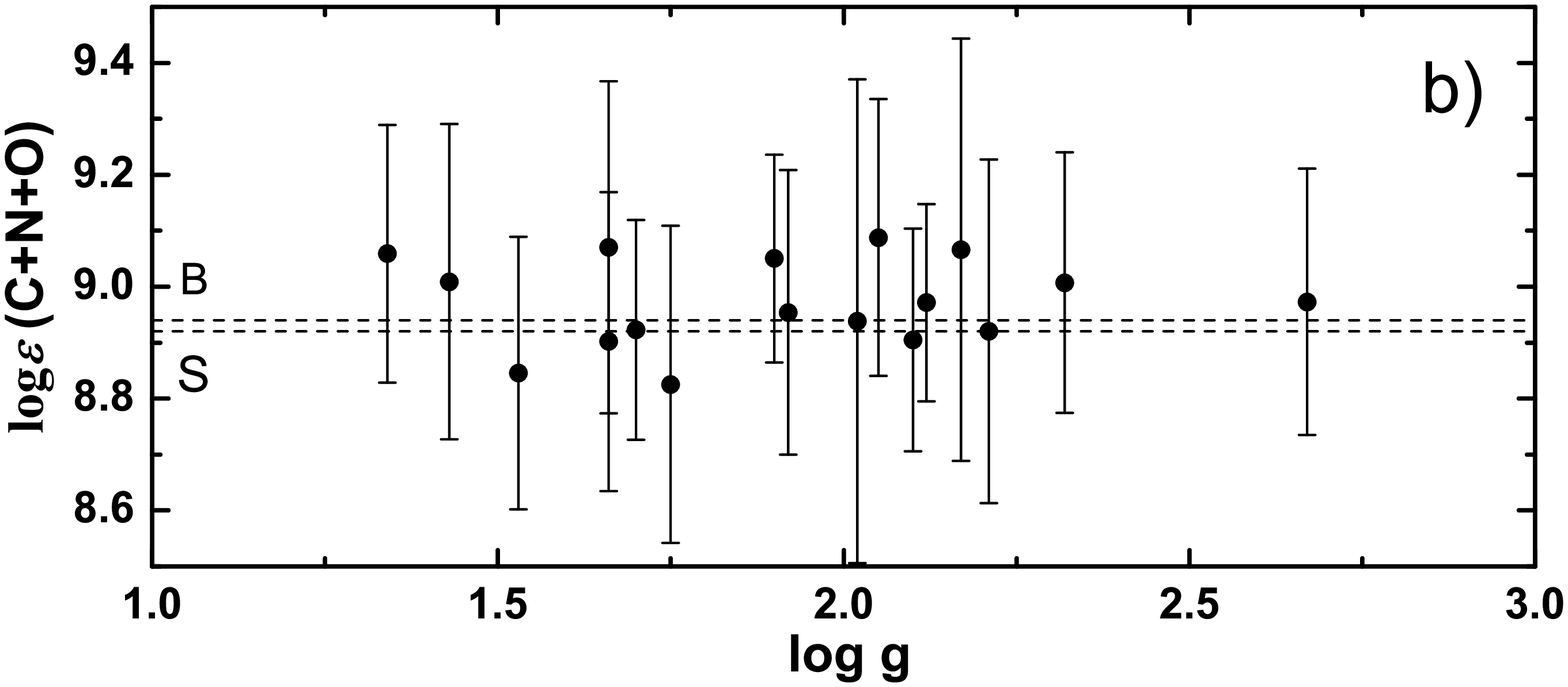}
\includegraphics[width=\columnwidth,clip=true]{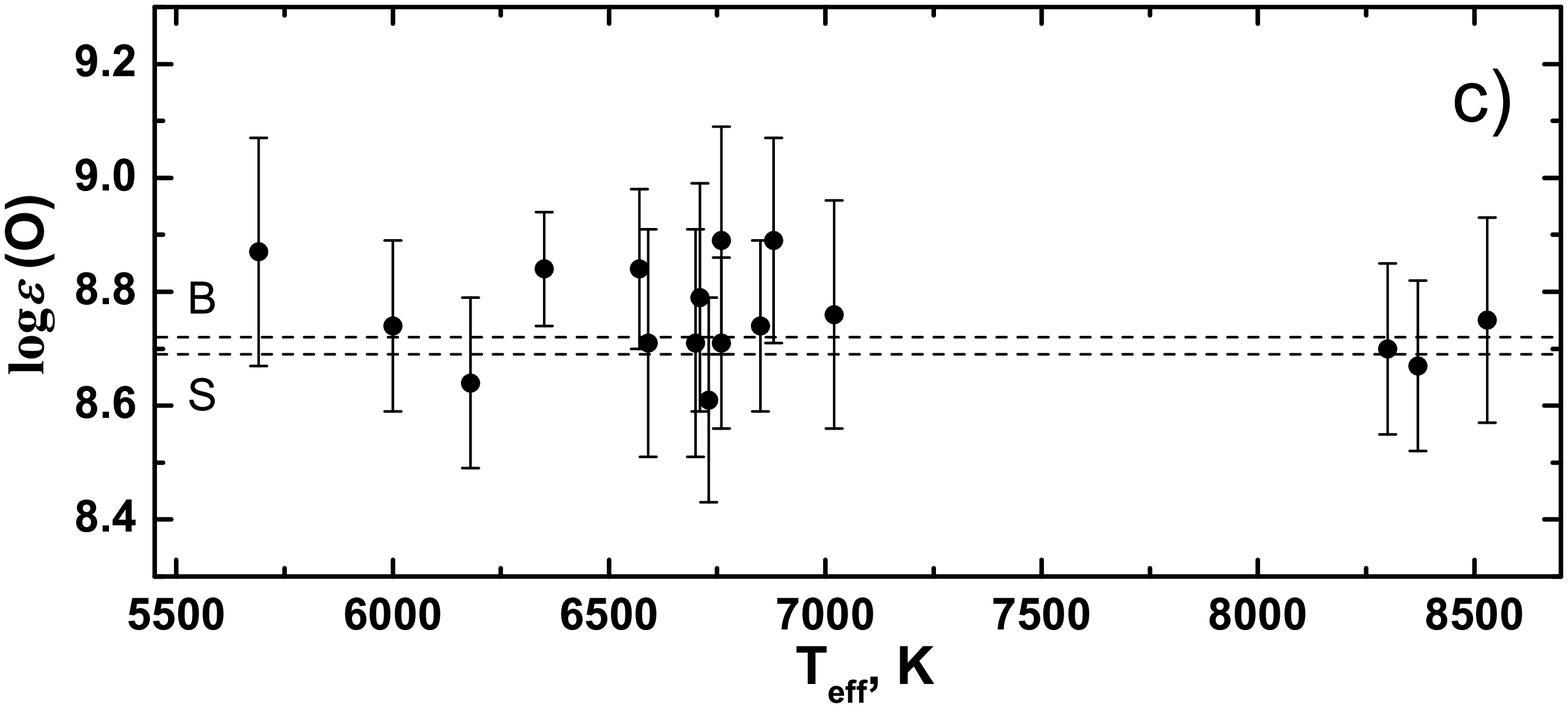}
\includegraphics[width=\columnwidth,clip=true]{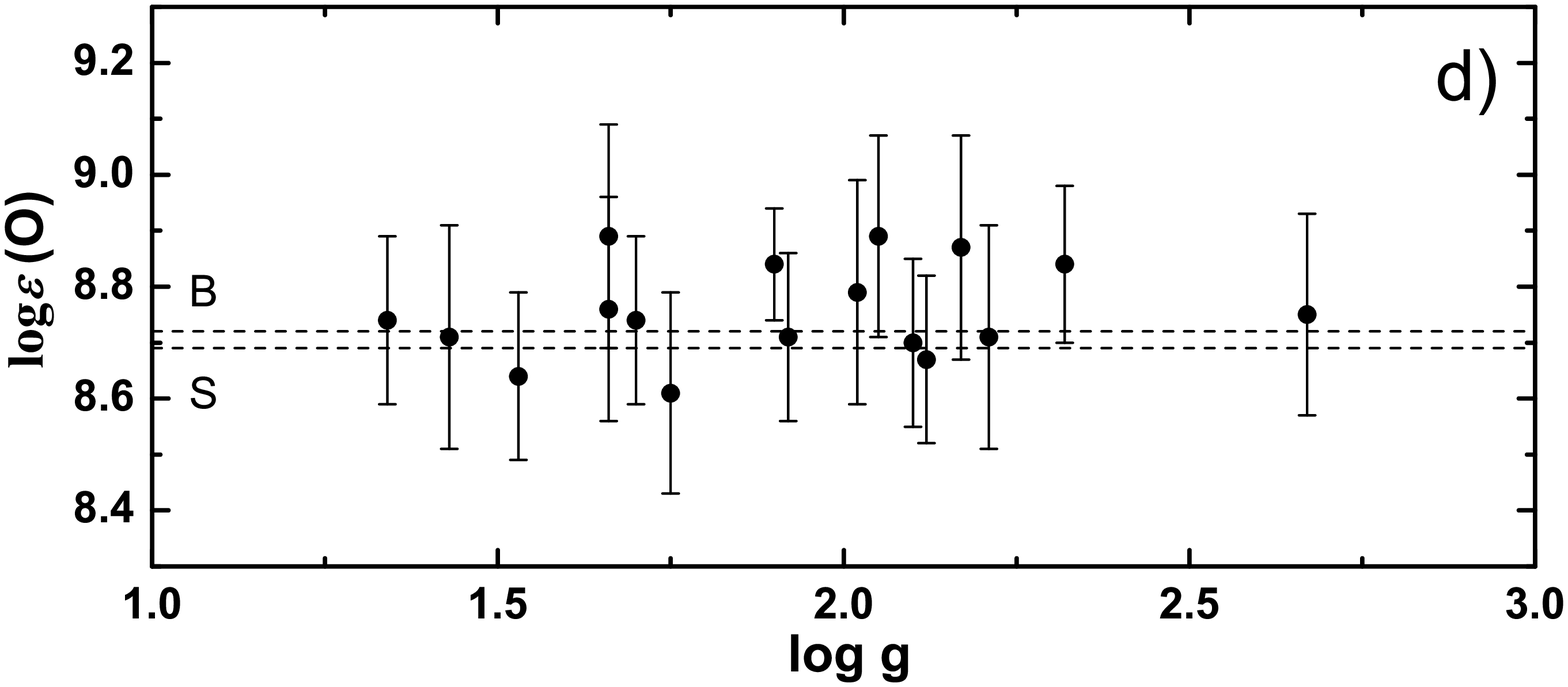}
\caption{\bf
The total abundance log $\varepsilon$(C+N+O) and the oxygen abundance 
log $\varepsilon$(O) for 17 stars from Table \ref{CNO} as functions of 
\teff\ and \logg. Dashed straight lines S and B correspond to the Sun 
\citep{Asplund09} and the unevolved early B-type stars \citep{Lyubimkov13}, 
respectively.}
\label{otl}
\end{figure}

The derived values log $\varepsilon$(C+N+O) for 17 stars are presented in 
last column of Table \ref{CNO}. Their dependences on the basic parameters 
\teff\ and \logg\ are shown in Fig. \ref{otl}a and Fig. \ref{otl}b, 
respectively. We show for comparison in Figs \ref{otl}c, d similar dependences 
for the oxygen abundances log $\varepsilon$(O) for the same 17 stars. One may 
see that there are no trends with \teff\ and \logg\ neither for 
log $\varepsilon$(C+N+O) nor for log $\varepsilon$(O). Moreover, there is a 
very good agreement (within error bars) with the initial values presented by 
dashed straight lines S (the Sun) and B (unevolved B stars). 

We obtained the mean value log $\varepsilon$(C+N+O) = 8.97$\pm$0.08 for 17 
stars in Fig. \ref{otl}, whereas the initial value is log $\varepsilon$(C+N+O) = 8.92 
(S) or 8.94 (B). Therefore, the sum C+N+O is actually conserved, and this fact 
confirms once more a good agreement between observations and the theory. 

Nevertheless, it should be stated that such accordance between observations and 
the theory exists not always. As noted in \citet{Lyubimkov18} review, there is a 
significant discrepancy between observations and the theory for helium, the 
main product of CNO-cycle. The same computations of \citet{Georgy13}, which 
seem to be successful in the case of C, N and O, cannot explain the observed 
helium enrichment in the B-type MS stars. The predicted He enrichment is too 
small, especially for the most typical initial velocities \v0 $<$ 150 km/s. It 
is evident that the theory should be substantially perfected in order to 
correspond to the observed He enrichment.

\section{Conclusions}

Thus, our non-LTE analysis of the oxygen abundance for 51 Galactic A-, F- and 
G-type supergiants and bright giants (luminosity classes I and II) led to the 
following conclusions. 

In contrast with carbon and nitrogen, oxygen did not show any significant 
systematic anomalies in their abundances log $\varepsilon$(O) for stars in question. There 
is no marked difference from the initial oxygen abundance within errors of the 
log $\varepsilon$(O) determination across the Teff interval from 4500 to 8500 K and the 
\logg\ interval from 1.2 to 2.9 dex. This result agrees well with theoretical 
predictions, which show for AFG supergiants in the post-FDU phase that the 
log $\varepsilon$(O) lowering is less than 0.15 dex if the initial rotational velocities 
\v0 $<$ 300 km/s. 

With our new results for oxygen and our earlier non-LTE determinations 
of the N and C abundances for stars from the same sample, we constructed the 
[N/C] vs [N/O] relation for 17 stars. A correlation between these values is 
found to be pronounced; the observed [N/C] increase from 0 to 1.6 dex is 
accompanied by the [N/O] increase from 0 to 0.9 dex. 

The observed [N/C] vs [N/O] relation is compared with the theoretical 
predictions for stellar models with rotation. Such a comparison shows that 
this relation reflects a strong dependence of the evolutionary changes in CNO 
abundances on the initial rotation velocities of stars. Early the same 
conclusion was made for the N vs C anticorrelation observed for these stars 
(Paper IV). 

It is shown that the stars studied are mostly the post-FDU objects with the 
initial rotational velocities \v0 $<$ 150 km/s. It is important that just such 
velocities \v0\ are typical for about 80 \% of stars in question, i.e. for stars 
with masses 4-19 M$_{\sun}$. It is confirmed that the star HR 7876 is the post-MS and 
the pre-FDU object with \v0 $\sim$ 100 km/s or less. The F supergiants HR 7387 and 
HR 1865 with the highest [N/C] values (1.21 and 1.56 respectively) are likely 
the post-FDU objects with \v0 $\approx$ 200-250 km/s and 400 km/s, respectively. 

A constancy of the total C+N+O abundance during stellar evolution is confirmed.
The mean value log $\varepsilon$(C+N+O) = 8.97$\pm$0.08 found by us for AFG 
supergiants and bright giants seems to be very close to the initial value 
8.92 (the Sun) or 8.94 (the unevolved B-type MS stars). 

Therefore, one may claim that theoretical predictions based on stellar models 
with rotations seem to be successful in explanation of the observed C, N and O 
abundances both in the B-type MS stars and in the AFG-supergiants and bright 
giants. However, it is necessary to note that the same models cannot explain 
the observed helium enrichment in the B-type MS stars. So, further perfecting 
of the theory is needed.

\section*{Acknowledgements}

We thank our referee for thorough comments and useful remarks.
DLL thanks the Robert A. Welch Foundation of Houston, Texas for support 
through grant F-634.

%%%%%%%%%%%%%%%%%%%%%%%%%%%%%%%%%%%%%%%%%%%%%%%%%%



\bibliographystyle{mnras}
\bibliography{O_I} % if your bibtex file is called example.bib

%\begin{thebibliography}{99}

%\end{thebibliography}

\label{lastpage}
\end{document}